\documentclass[twocolumn,showpacs,prd]{revtex4}
\usepackage{dcolumn}
\usepackage{bm}

\def\slashii#1{\setbox0=\hbox{$#1$}             
   \dimen0=\wd0                                 
   \setbox1=\hbox{\sl/} \dimen1=\wd1            
   \ifdim\dimen0>\dimen1                        
      \rlap{\hbox to \dimen0{\hfil\sl/\hfil}}   
      #1                                        
   \else                                        
      \rlap{\hbox to \dimen1{\hfil$#1$\hfil}}   
      \hbox{\sl/}                               
   \fi}                                         %
\def\slashiii#1{\setbox0=\hbox{$#1$}#1\hskip-\wd0\hbox to\wd0{\hss\sl/\/\hss}}
\def\slashiv#1{#1\llap{\sl/}}

\begin{document}

\title{Asymptotic Infrared Fractal Structure of the
Propagator for a Charged Fermion}

\author{S. Gulzari,
J. Swain and A. Widom}

\affiliation{Physics Department\\
Northeastern University \\
110 Forsyth Street, \\
Boston, MA 02115, USA\\
E-mail: john.swain@cern.ch}

\begin{abstract}
It is well known that the long-range nature of the Coulomb interaction makes 
the definition of asymptotic ``in'' and ``out'' states of charged particles
problematic in quantum field theory. In particular, the notion of a simple 
particle pole in the vacuum charged particle propagator is untenable and should 
be replaced by a more complicated branch cut structure describing an electron 
interacting with a possibly infinite number of soft photons. Previous work suggests 
a Dirac propagator raised to a fractional power dependent upon the fine structure 
constant, however the exponent has not been calculated in a unique gauge invariant 
manner. It has even been suggested that the fractal ``anomalous dimension'' can be 
removed by a gauge transformation. Here, a gauge invariant non-perturbative calculation 
will be discussed yielding an unambiguous fractional exponent. The closely analogous 
case of soft graviton exponents is also briefly explored.
\end{abstract}

\maketitle

\section{Introduction \label{intro}}

The propagator \begin{math} \tilde{S}(x-y) \end{math} for a free electron in a noninteracting 
theory describes the amplitude for an electron with a definite mass to propagate 
from {\it y} to {\it x}. For a theory with electromagnetic interactions, the picture is 
not so simple. The electron moving along its world line will be constantly interacting with its
surrounding cloud of virtual infrared photons. Truly free particles\cite{Weinberg1} are 
labeled by irreducible  representations of the Poincar\'{e} group in terms of mass and spin. 
However, charged particles cannot be consistently  assigned precise
masses\cite{Porrmann,Buchholz1,Buchholz2} since 
they continually interact with  massless photons. 
The resulting electromagnetic fields are too
long ranged to rigorously define ``in'' and ``out'' fixed mass states and one
must introduce concepts such as infraparticles (For a review see \cite{Schroer}).
In any case, additive perturbation theory is not sensible since it requires very  little 
energy to create and/or destroy an infinite number of photons  each with virtually 
zero energy. This invalidates computations which only include low orders in
\begin{math} \alpha = e^2/\hbar c  \end{math}.  It was shown long ago\cite{Dyson} that the power 
series expansion in integer powers of $\alpha$ for quantum electrodynamics cannot be convergent. 

There exist treatments of the photon cloud about a given electron in terms of coherent states 
including associated electromagnetic fields. The resulting wave functions lose the notion 
of a local single-particle electron concept\cite{Coherentstates}.
The wave functions describe the electromagnetic fields as superpositions of an infinite number 
of photon Fock states. The many particle (one electron plus infinitely many photons)
wave functions are quite complicated. In this work we seek a modified single electron
Dirac propagator which represents an electron with its associated infinite
number virtual photons without the need to explicitly employ the many
body wave functions in the final results.

There exist propagator calculations in the literature based on either infinite sums 
of logarithmic Feynman diagrams\cite{sumlogs} or non-perturbative 
Schwinger\cite{Schwinger,Fradkin,Rivers} computations which argue that such an electron
propagator should be of the form
\begin{equation}
S(k)=\left(\frac{\kappa}{i\Lambda}\right)^{\gamma }
\Gamma (1+\gamma ) \left\{\frac{\kappa -\slashiv{k}}
{\big[k^2+\kappa^2-i0^+\big]^{(1+\gamma )}}\right\},
\label{Appelquist}
\end{equation}
wherein \begin{math} \hbar \kappa =mc \end{math}, \begin{math} \Lambda  \end{math}
is a short distance length scale, the fractional exponent
\begin{math} \gamma \end{math} is a function of the coupling strength 
\begin{math} \alpha =(e^2/\hbar c) \end{math} and the Gamma function 
\begin{equation}
\Gamma (z)=\int_0^\infty e^{-s}s^z \frac{ds}{s}
\ \ \ {\rm with}\ \ \ {\Re e}(z) > 0.
\label{AppleG}
\end{equation} 
However, there is not yet any universal agreement about what constitutes the 
correct function \begin{math} \gamma (\alpha )\end{math}.

Appelquist and Carazzone\cite{sumlogsAC} take 
\begin{math} \gamma =-(\alpha/\pi )+\ldots \end{math} to leading order.
This differs from older work based on summing logarithms\cite{sumlogs}.
The fourth volume of the Landau and Lifschitz course of theoretical
physics\cite{landau} at first appears to be in agreement with
\begin{math} \gamma \neq 0 \end{math} but ultimately chooses \begin{math}
\gamma =0 \end{math} by   assigning a small mass to the photon. The
photon mass implies a broken gauge symmetry, as well as the broken
conformal invariance of the free Maxwell field.  The null result
\begin{math} \gamma =0  \end{math} is also in conflict with our physical
understanding that no sharp mass can be assigned to a charged  particle.
The effect of assigning the photon a small mass is clear from the 
physical picture of photon processes. As one backs away from the world
line of an  electron, thereby going to larger and larger wavelengths, a
photon mass vetoes  the production of more and more photons without limit
and the scaling behavior  of the electromagnetic fields is thereby broken
(see Supplement 4 of \cite{JauchandRohrlich}). The broken gauge symmetry
is  physically unacceptable. It has been argued\cite{Handel} that the
change from  a pole into a branch point has measurable physical
implications in measurements of  ``\begin{math} 1/\omega   \end{math}''
noise in the Schr\"{o}dinger non-relativistic  limit of the relativistic
Dirac equation. A path integral\cite{Rivers} approach  using the
Schwinger\cite{Schwinger} proper time representation of the propagator 
and some work by Bloch and Nordsieck\cite{BN} on soft photon emission
gives the  same sort of result, but with the final answer given only for
charged scalar fields,  and generally considered to be gauge invariant in
such a way that the singularity  structure can be returned to a simple
pole by a choice of gauge. Fried also  discusses this
problem\cite{Friedbook} as do Johnson and Zumino\cite{Johnson_Zumino},
and Stefanis and collaborators\cite{Stefanis}. Batalin, Fradkin and
Schvartsman have made a  similar gauge dependent calculation for scalar
particles\cite{Batalin}.

The major defects of the existing calculations are that they do not directly
give a Dirac propagator raised to a fractional power. Existing calculations 
are ambiguous and not gauge invariant. 

The fractional exponent result is rather dramatic. The fractional 
exponent implies some non-locality. The non-locality has been 
previously introduced ad-hoc\cite{nonlocalreg,Moffat} as a regularization tool. 
Here it appears naturally and also means that the singularity structure for 
the propagator is dramatically altered from a simple pole to a branch point. 
It also renders the analytic but rarely used regularization proposed by 
Speer\cite{Speer} more physically motivated.

\section{Gauge Invariant Calculation\label{GIA}}

For an electron moving through an external electromagnetic field, 
\begin{math} 
F_{\mu \nu}=\partial_\mu A_\nu -\partial_\nu A_\mu ,
\label{GIA1} 
\end{math}
the Dirac propagator 
\begin{eqnarray}
\left(-i\slashiii{d}+\kappa \right)G(x,y;A)=\delta (x-y)\ , 
\nonumber \\ 
d_\mu =\partial_\mu -i\left(\frac{eA_\mu }{\hbar c}\right)\ ,
\label{GIA2} 
\end{eqnarray}
may be solved employing the function 
\begin{math} \Delta (x,y;A) \end{math};  
\begin{eqnarray}
\Delta (x,y;A)=\int \gamma_5G(x,z;A)\gamma_5G(z,y;A)d^4z\ ,
\nonumber \\ 
G(x,y;A)=\left(i\slashiii{d}+\kappa \right)\Delta (x,y;A)\ ,
\nonumber \\  
\left(\slashiii{d}^2+\kappa^2 \right)\Delta (x,y;A) 
=\delta (x-y)\ ,
\nonumber \\
\slashiii{d}^2=-d^\mu d_\mu -
\frac{e}{2\hbar c}\sigma^{\mu \nu }F_{\mu \nu}\ .
\label{GIA3} 
\end{eqnarray}
Formally introducing the Hamiltonian of the electron as 
\begin{eqnarray}
{\cal H}_{tot}={\cal H}+{\cal H}_{spin}\ ,
\nonumber \\ 
{\cal H}=\frac{1}{2m}\left\{\left(p-\frac{e}{c}A\right)^2+m^2c^2\right\}\ ,
\nonumber \\ 
{\cal H}_{spin}=-\left(\frac{e\hbar }{4mc}\right)\sigma^{\mu \nu }F_{\mu \nu }
\label{GIA4}
\end{eqnarray}
and employing the operator representation \begin{math} p_\mu =-i\hbar \partial_\mu  \end{math}, 
one may define the amplitude for the electron to go from {\it y} to {\it x} in a proper time 
\begin{math} \tau \end{math} as the matrix element 
\begin{equation}
{\cal G}(x,y,\tau ;A)=\left<x\right|e^{-i{\cal H}_{tot}\tau /\hbar }\left|y\right>.
\label{GIA5}
\end{equation}
From Eqs.(\ref{GIA3}), (\ref{GIA4}) and (\ref{GIA5}) follows the electron propagator 
expression 
\begin{eqnarray}
\Delta (x,y;A)=\frac{i\hbar }{2m}\int_0^\infty 
{\cal G}(x,y,\tau ;A) d\tau ,
\nonumber \\ 
\hbar G(x,y;A)=\left(mc-\slashiv{p}+\frac{e}{c}\slashii{A}(x) \right)\Delta (x,y;A)\ .
\label{GIA6}
\end{eqnarray} 
The physical significance of \begin{math} {\cal H}(p,x) \end{math} can be made manifest 
in the formal classical limit \begin{math} \hbar \to 0  \end{math}. Hamilton's equations in 
proper time,
\begin{equation}
v^\mu =\frac{dx^\mu}{d\tau }=\frac{\partial {\cal H}}{\partial p_\mu }
\ \ \ {\rm and}\ \ \ f^\mu 
=\frac{dp_\mu }{d\tau }=-\frac{\partial {\cal H}}{\partial x^\mu },
\label{GIA7}
\end{equation} 
directly yield the the Lorentz force on a charge equation of motion 
\begin{equation}
m\frac{dv^\mu }{d\tau }=\frac{e}{c}F^{\mu \nu}v_\nu\ .
\label{GIA8}
\end{equation} 
Alternatively one may employ the Lagrangian formalism, 
\begin{eqnarray}
{\cal L}(v,x;A)=\frac{1}{2}m\left(v^\mu v_\mu -c^2\right)
+\frac{e}{c}v^\mu A_\mu (x),
\nonumber  \\ 
\frac{d}{d\tau}\left(\frac{\partial {\cal L}}{\partial v^\mu}\right)
=\left(\frac{\partial {\cal L}}{\partial x^\mu}\right);
\label{GIA9}
\end{eqnarray} 
i.e. Eqs.(\ref{GIA9}) also imply Eq.(\ref{GIA8}). 

For long wavelength modes of the electromagnetic field, one may with a sufficient 
degree of accuracy neglect the spin flip term in the Hamiltonian Eq.(\ref{GIA4}). 
In such a case we may approximate Eq.(\ref{GIA5}) by the Lagrangian 
{\em path integral} formulation   
\begin{eqnarray}
{\cal G}(x,y,\tau ;A)\approx \int_{X(0)=y}^{X(\tau )=x}
e^{i{\cal S}[X;A]/\hbar }\prod_\sigma dX(\sigma ), 
\nonumber \\ 
{\cal S}[X;A]=\int_0^\tau {\cal L}(\dot{X}(\sigma ),X(\sigma );A)d\sigma . 
\label{GIA10}
\end{eqnarray}  

Now, let us consider an electron following a world line path 
\begin{math} P \end{math} from {\it x} to {\it y} in a proper time 
\begin{math} \tau \end{math}. Since the {\em general} path   
\begin{math} P \end{math} contributing to the functional 
integral Eq.(\ref{GIA10}) represents the {\em virtual} motion of 
an electron, one finds, in general, that 
\begin{math} c^2\tau^2 \ne -(x-y)^2 \end{math}. It is only 
in the classical limit \begin{math} \hbar \to 0 \end{math} 
that \begin{math}  c^2d\tau^2 =-dX^\mu dX_\mu  \end{math}.
In quantum mechanics the amplitude for a process
is the coherent sum of all amplitudes for all the different ways 
in which that process could happen. Thus, one integrates through all possible
proper times the electron could accumulate while going from {\it x} to 
{\it y}. Consider two different paths, \begin{math} P_1 \end{math} and 
\begin{math} P_2 \end{math}, contributing to the the path integral in 
Eqs.(\ref{GIA10}). Although the endpoints {\it y} and {\it x} are the same, 
the proper time of the two paths are different. The relativistic ``toy 
model'' analogy is to consider two twins starting at the same age at  
{\it y} taking two different paths, \begin{math} P_1 \end{math} and 
\begin{math} P_2 \end{math}, and meeting again at {\it x} when their ages are 
in general different.

The interaction between the electron and the elecromagnetic vector 
potential is described by the action 
\begin{eqnarray}
S_{\rm int}(P;A)=\int_0^\tau {\cal L}_{\rm int}
(\dot{X}(\sigma ),X(\sigma );A)d\sigma ,
\nonumber \\ 
S_{\rm int}(P;A)=\frac{e}{c}\int_0^\tau A_\mu (X(\sigma ))
\dot{X}^\mu (\sigma ) d\sigma ,
\nonumber \\ 
S_{\rm int}(P;A)=\frac{e}{c}\int_P A_\mu (X) dX^\mu ,
\label{GIA11}
\end{eqnarray}
where the integral is along the world line \begin{math} P \end{math}.
To describe an electron moving through a vacuum region with zero point 
electromagnetic fields, one averages over the vacumm field fluctuations 
according to the rule 
\begin{eqnarray}
e^{iS_{\rm int}(P;A)/\hbar }\to 
\left<0\right|e^{iS_{\rm int}(P;\hat{A})/\hbar }\left|0\right>_+,
\nonumber \\  
e^{iS_{\rm int}(P;A)/\hbar }\to e^{iS_{self}(P)/\hbar },
\nonumber \\ 
S_{self}(P)=\frac{\hbar \alpha }{2}
\int_P \int_P D_{\mu \nu }(x_1-x_2)dx_1^\mu dx_2^\nu .
\label{GIA12}
\end{eqnarray} 
In the above Eq.(\ref{GIA12}), the subscript 
``\begin{math}+\end{math}'' denotes time odering, 
\begin{math} \hat{A}_\mu (x) \end{math} denotes the 
operator vector potential field and the photon propagator is given by 
\begin{equation}
D_{\mu \nu }(x_1-x_2)=\frac{i}{\hbar c}
\left<0\right|\hat{A}_\mu (x_1)\hat{A}_\nu(x_2)\left|0\right>_+.
\label{GIA13}
\end{equation}
The action form in Eq.(\ref{GIA12}) is of a well known form\cite{Feynman-Wheeler},  
and we have bypassed the usual Bloch-Nordsieck replacement of 
\begin{math} c\gamma^\mu \end{math} by 
four velocity \begin{math} v^\mu \end{math} by simply evaluating a phase 
in the soft photon infrared limit that we are considering.
The propagator may be written
\begin{eqnarray}
D_{\mu\nu}(x-y)=\left(\eta_{\mu \nu}-
(1-\xi)\frac{\partial_\mu \partial_\nu }{\partial^2}\right)D(x-y),
\nonumber \\ 
D(x-y)=\int \frac{4\pi }{k^2-i0^+}e^{ik\cdot (x-y)}\frac{d^4 k}{(2\pi )^4},
\nonumber \\ 
D(x-y)=\frac{i}{\pi }\left\{\frac{1}{(x-y)^2+i0^+}\right\}.
\label{GIA14}
\end{eqnarray}
where the parameter \begin{math} \xi \end{math} fixes a gauge.
Because the world line of the electron never begins nor ends 
(charge conservation), the partial derivative terms in 
Eq.(\ref{GIA14}) do not contribute to the self action in 
Eq.(\ref{GIA12}); Independently of the gauge paramemter 
\begin{math} \xi \end{math} we have 
\begin{equation}   
S_{self}(P)=\frac{\hbar \alpha }{2}
\int_P \int_P D(x_1-x_2){dx_1}_\mu dx_2^{\ \mu }.
\label{GIA15}
\end{equation} 

In the absence of any external field (above and beyond the vacuum 
fluctuation operator \begin{math} \hat{A} \end{math}) we have now 
derived expressions for the renormalized vacuum electron 
propator
\begin{eqnarray}
\tilde{G}(x-y)=\int S(k)e^{ik\cdot(x-y)}\frac{d^4k}{(2\pi )^4}\ ,
\nonumber \\ 
\tilde{G}(x-y)=\left<0\right|G(x,y;\hat{A})\left|0\right>_+\ ,
\nonumber \\ 
\tilde{G}(x-y)=\left(i\slashiii{\partial }+\kappa \right)
\tilde{\Delta }(x-y)\ ,
\nonumber \\ 
\tilde{\Delta }(x-y)=\frac{i\hbar }{2m}\int_0^\infty 
\tilde{\cal G}(x-y,\tau ) d\tau .
\label{GIA16}
\end{eqnarray}
The functional integral expression for 
\begin{math} \tilde{\cal G}(x-y,\tau ) \end{math}
is given by 
\begin{eqnarray}
\tilde{\cal G}(x-y,\tau )=\int_{X(0)=y}^{X(\tau )=x}
e^{i\tilde {\cal S}[X;A]/\hbar }\prod_\sigma dX(\sigma ), 
\nonumber \\ 
\tilde{\cal S}[X]=\int_0^\tau {\cal L}_0(\dot{X}(\sigma ))d\sigma 
+S_{self}[X] , 
\label{GIA17}
\end{eqnarray}  
wherein the {\em free} electron Lagrangian is 
\begin{equation}
{\cal L}_0(\dot{X})=\frac{1}{2}m_0(\dot{X}^\mu \dot{X}_\mu -c^2),
\label{GIA18}
\end{equation}
and the self action is diven by Eqs.(\ref{GIA14}) and (\ref{GIA15}) as 
\begin{equation}
S_{self}[X]=\frac{i\hbar \alpha }{2\pi }
\int_0^\tau \int_0^\tau 
\frac{\dot{X}^\mu (\sigma_1)\dot{X}_\mu (\sigma_2)d\sigma_1 d\sigma_2}
{(X(\sigma_1 )-X(\sigma_2))^2+i0^+}\ .
\label{GIA19}
\end{equation}
The divergent piece of the self action 
\begin{eqnarray}
{\Re }eS_{self}[X]=\frac{\Delta m}{2}\int_0^\tau 
(\dot{X}^\mu (\sigma )\dot{X}_\mu (\sigma )-c^2)d\sigma , 
\nonumber \\ 
|\Delta m|=\infty.
\label{GIA20}
\end{eqnarray}
The formally infinite {\em self mass} can be described by 
a finite physical mass \begin{math} 0<m=(m_0+\Delta m)<\infty \end{math}. 
Thus, Eq.(\ref{GIA17}) is renormalized to 
\begin{eqnarray}
\tilde{\cal G}(x-y,\tau )=\int_{X(0)=y}^{X(\tau )=x}
e^{i\tilde {\cal S}[X;A]/\hbar }\prod_\sigma dX(\sigma ), 
\nonumber \\ 
\tilde{\cal S}[X]=\int_0^\tau {\cal L}_m(\dot{X}(\sigma ))d\sigma 
+iW[X]
\nonumber \\  
{\cal L}_m(\dot{X})=\frac{1}{2}m(\dot{X}^\mu \dot{X}_\mu -c^2)
\nonumber \\  
W[X;\tau ]={\Im }mS_{self}[X],
\nonumber \\  
W[X;\tau ]=\frac{\hbar \alpha }{2\pi }
\int_0^\tau \int_0^\tau 
\frac{\dot{X}^\mu (\sigma_1)\dot{X}_\mu (\sigma_2)d\sigma_1 d\sigma_2}
{(X(\sigma_1 )-X(\sigma_2))^2}\ .
\label{GIA21}
\end{eqnarray}  
For a straight-line path \begin{math} X^\mu (\sigma )=V^\mu \sigma  \end{math} 
with \begin{math} V^\mu V_\mu =-c^2  \end{math}, one finds 
\begin{equation}
W(\tau )=\frac{\hbar \alpha }{2\pi }
\int_0^\tau \int_0^\tau \frac{d\sigma_1d\sigma_2}{(\sigma_1-\sigma_2)^2},
\label{GIA22}
\end{equation}
which can be made finite with the formal differential regularization\cite{Freedman}
\begin{equation}
\frac{d^2W(\tau )}{d\tau ^2}=\frac{\hbar \alpha }{\pi \tau^2}\ .
\label{GIA23}
\end{equation}
The solution to Eq.(\ref{GIA23}) with a logarithmic cut-off 
\begin{math} \Lambda  \end{math} is 
\begin{equation}
W(\tau )=-\left(\frac{\hbar \alpha }{\pi }\right)\ln \left(\frac{c\tau }{2\Lambda }\right).
\label{GIA24}
\end{equation}
From Eqs.(\ref{GIA21}) and (\ref{GIA24}), one finds 
\begin{equation}
\tilde{\cal G}(x-y,\tau )\approx e^{-W(\tau )/\hbar }
\tilde{\cal G}_m(x-y,\tau )
\label{GIA25}
\end{equation}
wherein \begin{math} \tilde{\cal G}_m(x-y,\tau ) \end{math} is the proper 
time Green's function for a particle of mass \begin{math} m \end{math}
with the corresponding free Lagrangian \begin{math} {\cal L}_m(\dot{X}) \end{math}. 
In detail and to exponentially lowest order in \begin{math} \alpha  \end{math}, 
\begin{eqnarray}
\tilde{\cal G}_m(x-y,\tau )=\int \left\{e^{-i\hbar (k^2+\kappa^2)\tau /2m}
e^{ik\cdot(x-y)} \right\}\frac{d^4k}{(2\pi )^4}\ ,
\nonumber \\ 
\tilde{\cal G}(x-y,\tau )=\left(\frac{c\tau }{2\Lambda}\right)^{\alpha /\pi }
\tilde{\cal G}_m(x-y,\tau ).\ \ 
\label{GAI26}
\end{eqnarray}
Eqs.(\ref{GIA16}) and (\ref{GAI26}) imply 
\begin{eqnarray}
\tilde{\Delta }(x-y)=
\int {\cal D}(k)e^{ik\cdot (x-y)}\frac{d^4k}{(2\pi)^4}\ ,
\nonumber \\ 
{\cal D}(k)=\left(\frac{\kappa}{i\Lambda}\right)^{\alpha /\pi }
\left\{\frac{\Gamma \big(1+(\alpha /\pi )\big)}
{\big[k^2+\kappa^2-i0^+\big]^{\big(1+(\alpha /\pi ) \big)}}\right\},
\nonumber \\ 
\tilde{G}(x-y)=\int S(k)e^{ik\cdot (x-y)}\frac{d^4k}{(2\pi)^4}\ ,
\nonumber \\ 
S(k)=(\kappa -\slashiv{k}){\cal D}(k)\ .
\label{GAI27}
\end{eqnarray}
Eq.(\ref{GAI27}) is equivalent to the central Eq.(\ref{Appelquist}) of 
this work with
\begin{equation}
\gamma =\frac{\alpha }{\pi } +\dots 
\label{GAI28}
\end{equation}
in agreement with the magnitude, but not the sign, of $\gamma$ in Appelquist and Carazzone
\cite{sumlogsAC}.

\section{Physical Intepretation \label{PI}}

It is interesting to consider what the physical interpretation of the non-integer
exponent in the radiatively corrected Dirac propagator means. First of all, the fact
that the exponent is non-integer means that the renormalized Dirac operator is  
non-local\cite{fraccalc}. This was, of course, to be expected since the 
electromagnetic field has infinite range.

One may also expect a degree of self-similarity at long wavelengths since the only
scale-breaking term in QED is the mass of the electron. Intuitively, one might  
think of stepping back farther and farther from an electron world line and seeing 
contributions to its dressed structure from longer and longer wavelengths, i.e.   
softer and softer virtual photons. This would suggest a fractal\cite{Mandelbrot} 
structure, which is made precise by the above derivation. Such notions of scaling 
and fractality are not new in QED and in quantum field theory in general, but are 
often considered in the high energy, ultraviolet limit\cite{Collins,GL}. In this 
case additional complications arise since more and more charged excitations must 
be included, but again, one sees fractional exponents in form of anomalous 
dimensions and the renormalization group -- another reflection of a non-trivially 
realized scale invariance in the theory, but now at short distances.

If the photon is given a mass, however small, this structure will break down 
asymptotically, since now there is a minimum energy required to create a virtual 
photon, and at distances greater than the corresponding Compton wavelength one will 
get the non-interacting Dirac propagator. This was done in by 
Lifshitz {\em et al.}\cite{landau}, and this argument makes clear how breaking gauge 
invariance, {\em i.e.} including a photon mass, removes the anomalous scaling behavior 
here derived for gauge invariant QED. 

The fact that a particle is non-localizable, at least in part due to its 
electromagnetic field which extends over all space, is interesting. The feeling  
of this calculation is such that at least part of what one thinks 
of as quantum-mechanical about an otherwise point-particle (its lack of 
localizability) may arise from the non-perturbative quantum mechanics of its 
self-interaction\cite{Hestenes}.

Finally, the path integral expressions for the QED charged Dirac propagator 
employed here allow one to assign a fractal dimension to the paths taken by charged
particles in the infrared limit. The fractal nature of particle
paths has been discussed in the literature (see, for
example \cite{FeynmanHibbs,HeyFractals}). 
Abbott and Wise\cite{AbbotandWise}, working from nonrelativistic quantum
mechanics, found 2 as the dimension of a
quantum mechanical path, as opposed to 1 for a classical path.
Cannata and Ferrari\cite{Cannata}
extended this work for spin-1/2 particles and find different results not only
in the classical and quantum mechanical limits, but also in the
non-relativistic and relativistic limits.
Intuitively one can understand the dimension 2 result for the nonrelativistic
quantum mechanical case by
thinking of the Schr\"{o}dinger equation as a diffusion equation in imaginary
time\cite{Nelson}. For diffusion one has the distance $r$ a particle covers in time $t$
satisfying a relationship of the form $t\propto r^d$ wherein $d$ is the fractal 
dimension of the ``path''. For example, $t\propto r^2$ in the diffusion limit,
and $t\propto r$ in the ballistic (simple path) limit\cite{Sornette}. 
Here we have a closely analogous situation but with a 4-dimensional Hamiltonian 
${\cal H}$  and with fractal diffusion in proper time. We find
\begin{equation}
d=2(1+\gamma )\approx 2+\frac{2\alpha }{\pi }+\ldots \ ,
\label{PI1}
\end{equation}
{\em i.e.} the ``path'' through space-time as a function of internal proper time 
has a dimension slightly higher than that of a two-dimensional surface. This
excess over dimension 2 can be thought of as due to an additional roughening
of the path of a charged
particle due to interactions with vacuum fluctuations.
All previous discussions have ignored the effects of self-interaction via
long-range fields.

\section{Extensions}

It is interesting to ask to what extent one might expect that similar behaviour
might occur with other interactions. With the weak interactions, the finite mass 
of the gauge bosons will cause this analysis to break down at distances of the 
order of their Compton wavelengths. The weak interactions are not infinite range.

For QCD, there is an additional problem in that the gluons, unlike the photons, now 
couple to each other, and with increasing strength at larger and larger distances,
so we do not expect the QED treatment to carry over very easily. That being said, one 
might well expect some sort of fractional exponent propagator with a  to appear. 
In the QED case it appears tfor large values of \begin{math} \alpha \end{math} 
radically different propagators would arise as the exponent hits zero and then changes sign. 
Some of this behaviour may be linked with confinement.

For quantum gravity one can do the analysis in much the same way as for 
quantum electrodynamics. The ultraviolet divergences to need not worry us  
since we are dealing with is a strictly infrared problem. There should be 
graviton-graviton interactions, but if we neglect these as 
small compared to graviton-electron interactions we can simply repeat what was done for 
electrostatics but now with the Newtonian limit of gravity. Note
that this sort of approximation could not be done 
consistently in the QCD case for a quark propagator since gluon-gluon couplings 
are comparable to gluon-quark couplings, but there is evidence from other
calculations of the appearance of anomalous dimensions for infrared propagators.
(For a review see \cite{Alkofer}. Very recent calculations 
can also be found in references \cite{Fischer}.)
Recognizing that we need to replace the repulsive electrostatic self-interaction,  
say \begin{math} +(e^2/r)  \end{math}, with the attractive 
gravitational self-interaction, say \begin{math} -Gm^2/r  \end{math},
suggests an asymptotic form of the Dirac propagator exponent    
\begin{equation}
\gamma \approx \frac{1}{\pi \hbar c }(e^2-Gm^2)+\ldots \ . 
\end{equation}
If \begin{math} m=|e/ \sqrt{G}|\end{math}, which is the ADM\cite{Ashtekar,ADM} 
mass of charged shell of charge \begin{math} e \end{math}, regularized by its own gravity, then one 
recovers an effectively free propagator. Given that one generally 
makes measurements using the electromagnetic interaction, and the suggestion that 
quantum mechanics might be linked to self-interaction\cite{Hestenes}, it is interesting 
to consider what this might imply for the role of gravity in the quantum measurement 
problem\cite{Penrose}. In particular, since one has a connection between mass 
and charge which is non-perturbative in Newton's \begin{math} G \end{math}, and implies 
a mass near the Planck mass \begin{math} \sim 10^{-5} \end{math} gm, which may be thought 
to be in the neighborhood of a putative classical-quantum boundary.

\section{Conclusions}

We have provided a simple and intuitive path integral description of how the 
propagator for a charged Dirac particle is modified by soft self-energy radiative 
corrections. The result is a self-similar (fractal) object with the non-locality one would 
expect for a particle carrying an infinite range field.
The extension of this 
treatment to other interactions was briefly explored, and the special case of soft graviton 
corrections quantitatively discussed. 
\medskip

\section{Acknowledgements}

We would like to thank Yogi Srivastava for useful conversations. We would also
like to thank Reinhard Alkofer for pointing out references \cite{JauchandRohrlich}
and \cite{Alkofer} to us, and Christian Fischer for pointing out \cite{Fischer}.
This work was
supported in part by a grant from the National Science Foundation NSF-0457001.

\end{document}